\documentclass[letter]{aa} 
\usepackage{graphicx}
\usepackage{color}

\usepackage{txfonts}
\usepackage{natbib}
\def \aap      {A\&A }

\def \apjl     {ApJL }
\def \apjs     {ApJS }

\def\gr{$\gamma$-ray}

\begin{document}
   \title{Measuring the correlation length of intergalactic magnetic fields from observations of gamma-ray induced cascades}

    \author{          
          A.Neronov\inst{1},
          A.M.Taylor\inst{2},
          C.Tchernin\inst{1}
          and Ie. Vovk\inst{1}
          }

\institute{{ISDC Data Centre for Astrophysics, Department of Astronomy, University of Geneva, Ch. d'Ecogia 16, CH-1290, Versoix, Switzerland},\\
\vspace{-2mm}
\and
{Dublin Institute for Advanced Studies, 31 Fitzwilliam Place, Dublin 2, Ireland}}


 
\vspace{-2mm}
  \abstract
   {The imaging and timing properties of \gr\ emission from electromagnetic cascades initiated by very-high-energy (VHE) \gr s in the intergalactic medium depend on the strength $B$ and correlation length $\lambda_B$ of intergalactic magnetic fields (IGMF). }
   {We study the possibility of measuring both $B$ and $\lambda_B$ via observations of the cascade emission with \gr\ telescopes.  }
   {For each measurement method, we find two characteristics of the cascade signal, which are sensitive to the IGMF $B$ and $\lambda_B$ values in different combinations. For the case of IGMF measurement using the observation of extended emission around extragalactic VHE \gr\ sources, the two characteristics are the slope of the surface brightness profile and the overall size of the cascade source. For the case of IGMF measurement from the time delayed emission, these two characteristics are the initial slope  of the cascade emission light curve and the overall duration of the cascade signal.}
   {We show that measurement of the slope of the cascade induced extended emission and/or light curve can both potentially provide measure of the IGMF correlation length, provided it lies within the range 10~kpc$\lesssim\lambda_B\lesssim$1 Mpc.  For correlation lengths outside this range, gamma-ray observations can provide upper or lower bound on $\lambda_B$. The latter of the two methods holds great promise in the near future for providing a measurement/constraint using measurements from present/next-generation \gr -telescopes.}
   {Measurement of the IGMF correlation length will provide an important constraint on its origin. In particular, it will enable to distinguish between an IGMF of galactic wind origin from an IGMF of cosmological origin.}


   \maketitle

\section{Introduction}

\vspace{-2mm}
Observations of the absorbed \gr\ component from high energy blazars have in recent times been used to provide constraints on the physical parameters of the IGM such as the density of optical/infrared radiation backgrounds \citep{franceschini08,Orr_EBL_1ES0229}. Furthermore, the subsequent electromagnetic cascade produced in the IGM \citep{aharonian94} provides the opportunity to probe the intergalactic magnetic field (IGMF) strength \citep{plaga95,neronov07,ichiki08,murase08,takahashi08}. Recent negative results on the search of the secondary \gr\ emission from the \gr\ induced electromagnetic cascades in the GeV energy band have been used to derive lower bounds on the IGMF strength and correlation length \citep{neronov_vovk,tavecchio10,dermer11,taylor11,vovk12}. Furthermore, such fields must necessarily permeate a significant fraction ($>$60\%) of the column depth to the blazar in order to exert a  sufficient effect on the electromagnetic cascade development \citep{dolag10}.
Thus, combining the obtained lower bounds with the known upper bounds from radio data and theoretical considerations, one finds that the allowed range of IGMF parameters spans several decades in both strength ($10^{-17}<B<10^{-9}$~G) and correlation length ($10^{13}<\lambda_B<10^{28}$~cm) parameter space. 

The allowed region of $(B,\lambda_B)$ parameter space is consistent with various intergalactic magnetic field generation scenarios, from phase transitions in the Early Universe \citep{hogan83, quashnock89, vachaspati91, sigl97} to supernova and AGN generated outflows from the galaxies during the recent Cosmological epoch \citep{bertone}. In general, relic cosmological magnetic fields from phase transitions in the Early Universe are expected to have short correlation lengths which depend on the magnetic field strength 
\vspace{-1mm}
\begin{equation}
  \lambda_{B,\rm PT}\sim 50\left[\frac{B}{10^{-9}\mbox{ G}}\right]\mbox{ kpc}
\end{equation}
which corresponds to the largest processed eddy scale-length at the end of the radiation dominated epoch \citep{jedamzik}. 
An exception to this rule are magnetic fields produced during the epoch of Inflation. In this case the initial correlation length of magnetic field could be arbitrarily large, so that the correlation length of the remaining magnetic field at zero cosmological redshift might also be arbitrarily large. However, the largest processed scale-length still limits the correlation length of the relic inflationary magnetic field to be $\lambda_{B,\rm Inflation}\gtrsim 50\left[\frac{B}{10^{-9}\mbox{ G}}\right]\mbox{ kpc}$.

In comparison, magnetic fields generated by galaxy outflows at the late stages of the Universe's evolution are expected to have correlation lengths of the order of typical galaxy sizes 
\vspace{-1mm}
\begin{equation}
  10\mbox{ kpc}\lesssim\lambda_{\rm B, gal}\lesssim 100\mbox{ kpc}
\end{equation}

Different possible scenarios for the generation of intergalactic magnetic field can potentially be distinguished through the measurement of $\lambda_B$. In what follows we show that gamma-ray measurements of emission from electromagnetic cascades in the intergalactic medium can provide a measurement of magnetic field correlation length if it lies within the range 10~kpc$\lesssim \lambda_B\lesssim 1$~Mpc and provide upper or lower bounds on $\lambda_B$ outside this range for all measurable values of magnetic field strength $B$. We note that the effects of plasma physics on cascade development within the voids, the importance of which remains unresolved \citep{broderick12,schlickeiser12,miniati12}, are neglected in this work.

\section{IGMF coherence length from the source angular profile}
\label{sect::angular_profile}

\begin{figure}
  \includegraphics[width=\columnwidth,angle=0]{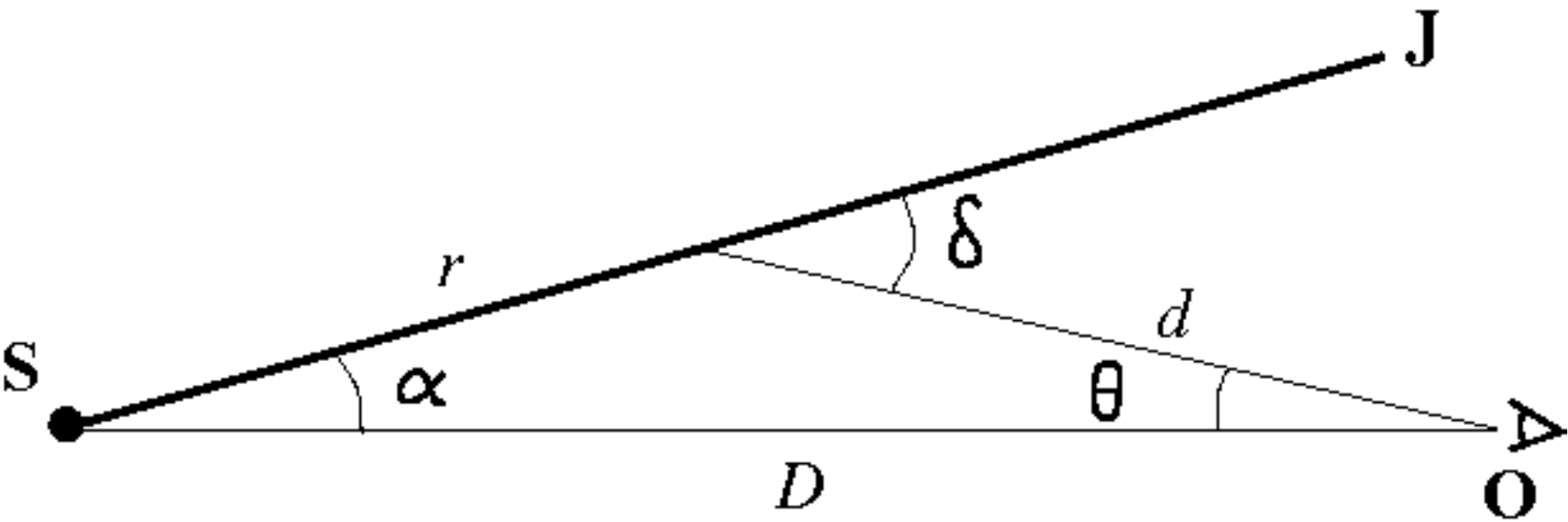}
  \caption{Geometry of cascade emission from a blazar jet SJ misaligned with the line of sight SO.}
  \label{fig::scheme1}
\end{figure}

{\bf Pencil Beam Model-} In order to derive simplified analytic expressions for cascade quantities, we start with the two-generation model depicted in Fig. \ref{fig::scheme1}.
Considering a narrow beam of \gr s emitted by a source S in a direction SJ misaligned with the line of sight SO (Fig. \ref{fig::scheme1}). For simplicity, we suppose that the primary VHE \gr\ beam consists of photons with the same energy $E_{\gamma_0}$ emitted by the source S at constant rate of  $N_0$ \gr s per unit time. 

Interaction of the VHE \gr s with extragalactic background light (EBL) leads to their absorption. As a result, the number of photons along the primary \gr\ beam decreases with distance as $N(r)=N_0\exp\left(-r/D_{\gamma_0}\right)$, where $D_{\gamma_0}$ is the mean free path of \gr s through the EBL. Each primary photon produces two electrons, so that the rate of injection of $e^+e^-$ pairs along the \gr\ beam is 
\begin{equation}
  \frac{dN_e}{dr}=\frac{2N_0}{D_{\gamma_0}}\exp\left(-\frac{r}{D_{\gamma_0}}\right)
\end{equation}
The subsequent cooling of these $e^+e^-$ pairs due to the Inverse Compton (IC) scattering of the cosmic microwave background (CMB) photons leads to emission of the secondary $\gamma$-ray photons with energies $E_\gamma\simeq 1 \left( E_{\gamma_0}/1\mbox{ TeV} \right) \mbox{ GeV}$. We suppose that the corresponding cooling distance
\begin{equation}
  D_e\simeq 0.2\left( \frac{E_\gamma}{1\mbox{ GeV}} \right)^{-1/2}\mbox{ Mpc}
\label{de}
\end{equation}
is much smaller than distance to the source $D$. This means that the energy absorbed from the initial beam is emitted \textit{in situ}, immediately after absorption.

The conservation of energy demands that the total power of IC emission from the $e^+e^-$ pairs is equal to the power removed from the primary \gr\ beam:
\begin{equation}
  \frac{dP_{IC}}{dr}=\frac{E_{\gamma_0} N_0}{D_{\gamma_0}}\exp\left(-\frac{r}{D_{\gamma_0}}\right)
  \label{eq::IC_power}
\end{equation}

In the presence of IGMF, electrons and positrons are deflected from their original directions. The deflection angle of these pairs depends on the correlation length of magnetic field, $\lambda_B$. Two different deflection regimes can be identified:\\


\noindent \textit{\bf ``one cell'' regime-} ($D_e\ll \lambda_B$). In this case, the $e^+ e^-$ pairs move in nearly homogeneous magnetic field. The deflection angle changes as
\begin{equation}
  \delta(x)=\frac{x}{R_L}
\end{equation}
where $x$ is the coordinate along the electron trajectory counted from the pair production point and $R_L=E_e/eB$ is the Larmor radius. As the characteristic value of $x$ is $D_e$, the previous equation can be rewritten as: 
\begin{equation}
  \delta=\frac{D_e}{R_L} \simeq 3 \times 10^{-4} \left( \frac{B}{10^{-16}~{\rm G}} \right)   \left( \frac{E_e}{10~{\rm TeV}} \right)^{-2}, 
\end{equation}

\noindent \textit{\bf ``many cells'' regime-} ($D_e\gg\lambda_B$). In this case particles move through many regions with different orientations of magnetic field. Under such conditions electrons and positrons experience random walks in angle, so that the average deflection angle is given by:
\begin{equation}
\delta(x)=\frac{\sqrt{\lambda_B x}}{R_L}
\end{equation}
or, substituting $x=D_e$:
\begin{equation}
\delta=\frac{\sqrt{\lambda_B D_e}}{R_L} \simeq 5 \times 10^{-5}  \left( \frac{E_e}{10~{\rm TeV}} \right)^{-3/2}  \left( \frac{B}{10^{-16}~{\rm G}} \right)  \left( \frac{\lambda_B}{1 ~{\rm kpc}} \right)^{1/2}
\end{equation}


The deflections of electrons\footnote{For the simplicity of the argument we will refer to both electrons and positrons as ``electrons''} by the IGMF determine the angular pattern of the IC emission. The angular distribution of electrons at each point along the beam can be written in the following way: 

\begin{equation}
\frac{\partial^2 N_e(\delta)}{\partial r \partial \Omega} = 
 \frac{\partial N_e(\delta)/\partial r}{2\pi\delta \partial \delta} =
 \frac{d N_e/dr \cdot f_\delta d\delta}{2\pi\delta d\delta} 
 \label{Eq:Angular_spread_1}
\end{equation}

The factor $f_\delta d\delta$ here denotes the fraction of the total number of particles, that were deflected within the range $[\delta : \delta+d\delta]$, and $N_e(\delta)$ is the corresponding number of electrons. We assume here, that the probability density function $f_x$ for the electron to emit secondary $\gamma$ ray is constant over its trajectory and drops to zero at $x=D_e$. The dependence of this density function on the deflection angle can then be written in the following way:
\begin{equation}
 f_\delta d\delta = f_x dx = f_x \frac{dx}{d\delta} d\delta = const \frac{dx}{d\delta} d\delta
\end{equation}
The constant here is defined through the requirement that $\int_{0}^{D_e} f_x dx = 1$.
We thus can rewrite Eq.~\ref{Eq:Angular_spread_1} in the following way:
\begin{eqnarray}
   \frac{\partial^2 N_e(\delta)}{\partial r \partial \Omega} \sim 
   \frac{dN_e/dr \cdot \frac{dx}{d\delta} d\delta}{2\pi\delta d\delta} = 
   \frac{dN_e/dr}{2\pi\delta (d\delta/dx)} = \\
   = \frac{dN_e/dr}{2\pi c}   \left\{
   \begin{array}{ll}
      R_L/\delta,& \lambda_B\gg D_e\\
      2R_L^2/\lambda_B,& \lambda_B\ll D_e
   \end{array}
   \right. \nonumber
\end{eqnarray}



The energy of electrons also changes with the distance $x$ as $E_e(x)=E_0/(1+x/D_e)$, where $E_0\simeq E_{\gamma_0}/2$ is the initial energy at the injection point. As the power of the IC emission scales as $E_e^2$, we can write:
\begin{eqnarray}
  \label{eq::fic}
  \frac{\partial^2 P_{IC}(\delta)}{\partial r \partial \Omega} & \sim & \frac{\partial^2 N_e(\delta)}{\partial r \partial \Omega} E_e^2(x[\delta]) \nonumber \\
  \frac{\partial^2 P_{IC}(\delta)}{\partial r \partial \Omega} & = &
  \frac{E_{\gamma_0}N_0R_L}{2\pi D_eD_{\gamma_0}} \exp\left(-\frac{r}{D_{\gamma_0}}\right) \times \\
  & \times & \left\{
  \begin{array}{ll}
     \frac{\displaystyle 1}{\displaystyle\delta\left(1+R_L\delta/D_e\right)^2} ,& \lambda_B\gg D_e\\
     \\
     \frac{\displaystyle 2R_L}{\displaystyle \lambda_B}\frac{\displaystyle 1}{\displaystyle\left(1+R_L^2\delta^2/(D_e\lambda_B)\right)^2} ,& \lambda_B\ll D_e
  \end{array}
  \right.\nonumber
\end{eqnarray}
where the normalization of $\frac{\partial^2 P_{IC}(\delta)}{\partial r \partial \Omega}$ is fixed in such a way that $\int_0^{\delta_{max}}  \frac{\partial^2 P_{IC}(\delta)}{\partial r \partial \Omega} 2\pi \delta d\delta=dP_{IC}/dr$ with $\delta_{max}=D_e/R_L$ in the case $\lambda_B\gg D_e$ and $\delta_{max}=\sqrt{\lambda_B D_e}/R_L$ in the case $\lambda_B\ll D_e$. 

An observer looking at the jet SJ with angle $\alpha$ will be able to observe the cascade emission from the $e^+e^-$-pairs deposited along the jet as long as the off-axis angle of IC \gr s emitted in the direction of the observer is $\delta<\delta_{max}$ (see Fig. \ref{fig::scheme1}). The flux of IC emission detected by an observer at point O depends on the angular distance from the source $\theta$. Taking into account that  $r=D\sin\theta/\sin\delta$ (Fig. \ref{fig::scheme1}), one can calculate the jet brightness profile

\begin{eqnarray}
  \label{eq:linear}
  \frac{dF}{d\theta} & = & \frac{dF}{dr}\frac{dr}{d\theta} = \frac{1}{d^2} \frac{\partial^2 P_{IC}(\delta)}{\partial r \partial \Omega} \frac{dr}{d\theta}
  = \frac{E_{\gamma_0}N_0R_L}{2\pi \alpha D^2D_eD_{\gamma_0}}\\
  & \times & \exp\left(-\frac{D\theta}{\delta D_{\gamma_0}}\right)\left\{
  \begin{array}{ll}
    \frac{\displaystyle 1}{\displaystyle \delta\left(1+R_L\delta/D_e\right)^2},& \lambda_B\gg D_e\\
    &\\
    \frac{\displaystyle 2R_L}{\displaystyle\lambda_B}\frac{\displaystyle1}{\displaystyle\left(1+R_L^2\delta^2/(D_e\lambda_B)\right)^2},& \lambda_B\ll D_e
  \end{array}
  \right.
  \nonumber
\end{eqnarray}

where $\delta=\alpha+\theta$. 


{\bf Jet Openining Angle Effects-} If the blazar jet is aligned to the line of sight, the cascade emission appears as an extended "halo-like" emission around the primary \gr\ source, rather than as a one-sided jet-like extension. In this case the measurable characteristic of the cascade emission is the slope of the cascade source's surface brightness profile (rather than the linear brightness profile of the jet-like extension).  

The cascade emission surface brightness profile of a \gr\ beam with an opening angle $\alpha_{jet}$ aligned along the line of sight can be found by summing the linear profiles (\ref{eq:linear}) of all the narrow \gr\ beams forming the jet: 
\begin{equation}
  \frac{d{\cal F}}{d\theta}=\frac{1}{2\pi\theta}\int_{\alpha_{min}}^{\alpha_{jet}}2\pi\alpha d\alpha\frac{dF}{d\theta}
  \label{eq::integral_angular_profile}
\end{equation}
where $\alpha_{min}$ is determined by the condition $\alpha_{min}=\theta(\tau-1)$, which can be understood from Fig.~\ref{fig::scheme1}. Indeed, in our calculations we assume that the characteristic distance $r$, at which $e^+e^-$-pairs are produced, is $D_{\gamma 0}$. Making this substitution in the limit of small $\alpha$ and $\theta$ one finds a condition $\alpha_{min}/(D-D_{\gamma 0}) = \theta/D_{\gamma 0}$, which then transforms in the above lower bound of the integral in Eq.~\ref{eq::integral_angular_profile}. Substituting $dF/d\theta$ from (\ref{eq:linear}) and taking the integral one finds in the limit of small $\theta$
\begin{equation}
\frac{d{\cal F}}{d\theta}\sim 
 \left\{
\begin{array}{ll}
\theta^{-1}(1+\mbox{ln}(\tau\theta)),& \lambda_B\gg D_e\\
&\\
\mbox{ const},& \lambda_B\ll D_e
\end{array}
\right.
\nonumber
\label{eq:linear1}
\end{equation}
Thus, in the case $\lambda_B\ll D_e$, the  cascade emission is disk-like with a flat surface brightness profile. To the contrary, in the case $\lambda_B\gg D_e$, the cascade emission has a steep brightness profile peaked at the central source. 

The dependence of the extended emission's surface brightness profile on $\lambda_B$ can therefore potentially be used for the measurement of the IGMF correlation length $\lambda_B$. Indeed, measurement of a non-zero slope in the surface brightness profile at energy $E_\gamma$ would imply the constraint  $\lambda_B>D_e$. To the contrary, measurement of a flat profile  would impose the constraint $\lambda_B<D_e$.  If $\lambda_B$ is larger than the IC cooling distance of the highest energy electrons, but shorter than the cooling distance of the lowest energy electrons contributing to the cascade \gr\ emission detectable by a \gr\ telescope, one may hope to detect a change in the slope of the brightness profile of the cascade emission at the energy $E_{\gamma,br}$ where $D_e\sim \lambda_B$. In this case, measurement of the break energy $E_{\gamma,br}$ would provide a measurement of the IGMF correlation length
\begin{equation}
  \lambda_B = 0.2 \left( \frac{E_{\gamma,br}}{1\mbox{ GeV}} \right)^{-1/2}\mbox{ Mpc}
  \label{eq::lambda1}
\end{equation}
It should be noted, however, that a measure of the coherence length employing this method
would require considerable improvement in angular resolution 
relative to that of the $\sim 0.1^{\circ}$ present day limit for \gr\ telescopes. In
the following section we describe an alternative method for measuring the
coherence length, potentially employable using present/next generation \gr\ instruments.

\section{IGMF coherence length from the source flare light curve}

A different regime of deflection of electrons by IGMF affects not only the slope of the profile of cascade emission but also the time delay of the cascade photons. Calculation of the temporal characteristics of the cascade emission signal can be done in a similar way to the calculation of the linear and/or surface brightness profiles, performed in the previous section. As was done there, we consider two model situations: a narrow jet misaligned with the line of sight and a finite opening angle jet aligned with the line of sight. 


{\bf Pencil Beam Model-} We consider again a jet SJ misaligned by an angle $\alpha$ with respect to the line of sight SO (Fig. \ref{fig::scheme1}). However, instead of constant in time injection of the primary \gr s at the source, we consider an instantaneous injection of $N_0$ \gr s. The \gr s propagate along the jet and deposit $dN_e/dr$ pairs on the time scale of the light crossing time of the distance  $r$. At any given time, only electrons injected in the cascade over the time interval $D_e/c$ contribute to the IC radiation in the cascade, so that the total number of the highest energy electrons is $N_e(r)=D_edN_e/dr$. The portion of the \gr\ beam which produces cascade emission detectable with the time delay $t_d$ compared to the direct signal from the source is situated at distance 
\begin{equation}
r=D\left(1-\frac{D\alpha^2}{(D\alpha^2+2ct_d)}\right)
\end{equation}
on the SJ line.
Following the calculation in section~\ref{sect::angular_profile}, the amount primary photon energy converted into the cascade \gr\ emission within distance interval $dr$ and angular distribution of the cascade emission are given by Eq.~\ref{eq::IC_power} and~\ref{eq::fic}, so we can now write the amount of the cascade emission which reaches the observer per time interval $dt_d$ as
\begin{eqnarray}
  \label{eq::Fict}
  \frac{dF_{IC}}{dt_d} & = & \frac{1}{d^2} \frac{\partial^2 P_{IC}(\delta)}{\partial r \partial \Omega} \frac{dr}{dt_d} = \frac{cE_{\gamma_0}N_0R_L}{\pi \alpha^2 D^3D_{\gamma_0}}
  \\ 
  & \times & \exp\left(-\frac{r}{D_{\gamma_0}}\right)
  \left\{
  \begin{array}{ll}
    \frac{\displaystyle 1}{\displaystyle\delta\left(1+R_L\delta/D_e\right)^2},& \lambda_B\gg D_e\\
    &\\
    \frac{\displaystyle 2R_L}{\displaystyle \lambda_B}\frac{\displaystyle 1}{\displaystyle\left(1+R_L^2\delta^2/(D_e\lambda_B)\right)^2},& \lambda_B\ll D_e
  \end{array}
  \right.\nonumber
\end{eqnarray}

where $r$ and $\delta=\alpha+2ct_d/D\alpha$ are expressed as functions of $t_d$.


{\bf Jet Opening Angle Effects-} The light curve of cascade emission from a jet of finite opening angle aligned with the line of sight can be obtained by summation of the light curves of all the beams forming the jet, i.e. via integration over the angle of the beam with respect to the line of sight $\alpha$:
\begin{equation}
  \frac{d{\cal F}_{IC}}{dt_d}=\int_{\alpha_{min}}^{\alpha_{jet}}2\pi\alpha d\alpha \frac{dF_{IC}}{dt_d}
  \label{eq::integral}
\end{equation}
where $\alpha_{jet}$ is the jet opening angle and $\alpha_{min}$  is found from the condition $D/\sin\delta=D_{\gamma_0}/\sin\alpha_{min}$, which gives 
\begin{equation}
  \alpha_{min}=\sqrt{\frac{2(\tau-1)ct_d}{D} }
\end{equation}
Substituting the expression for $dF_{IC}/dt_d$ (\ref{eq::Fict}) into (\ref{eq::integral}) and taking the integral one finds at the limit of small $t_d$:
\begin{equation}
  \label{eq::analytic}
  \frac{{d\cal F}_{IC}}{dt_d}\sim\left\{
  \begin{array}{ll}
    t_d^{-1/2},& \lambda_B\gg D_e\\
    const,& \lambda_B\ll D_e
  \end{array}
  \right.
\end{equation}

The slope of the cascade emission light curve depends on the relation between the IGMF correlation length and electron cooling distance. This fact can be used for the measurement of $\lambda_B$. At a fixed cascade photon energy, measurement of a flat cascade emission light curve would impose an upper bound on the IGMF correlation length, $\lambda_B\ll D_e$. For the opposite case, a lower bound on $\lambda_B$ would be set.
If $\lambda_B$ is larger than the IC cooling distance of the highest energy electrons, but shorter than the cooling distance of the lowest energy electrons contributing to the cascade \gr\ emission detectable by a \gr\ telescope, one expects to find a change in the slope of the  cascade emission light curve at the energy $E_{\gamma,br}$ where $D_e\sim \lambda_B$. In this case, measurement of the break energy $E_{\gamma,br}$ would provide a measurement of the IGMF correlation length, given by Eq.~\ref{eq::lambda1}. The range of length scales probable by this method is therefore dictated by the dynamic energy range of the instrument. For the example case of Fermi LAT, with a dynamic range of approximately 4 decades (20~MeV to 300~GeV \citep{2009ApJ...697.1071A}), the corresponding coherence scale range probable by this method is 10~kpc$\lesssim \lambda_B\lesssim 1$~Mpc.


\section{Verification with Monte Carlo Simulation}

Since relations (\ref{eq:linear1}) and (\ref{eq::analytic}) have been obtained 
using a simplified two-generation model, we here compare these results against
those obtained with a complete multi-generation numerical description. 
Using the Monte Carlo simulation described in \cite{taylor11} in which
the full cascade development is carried out and the spatial deflection
of the electrons tracked a comparison of the two-generation results was
carried out. 

\begin{figure}[ht]
\begin{center}
\includegraphics[width=\columnwidth,angle=0]{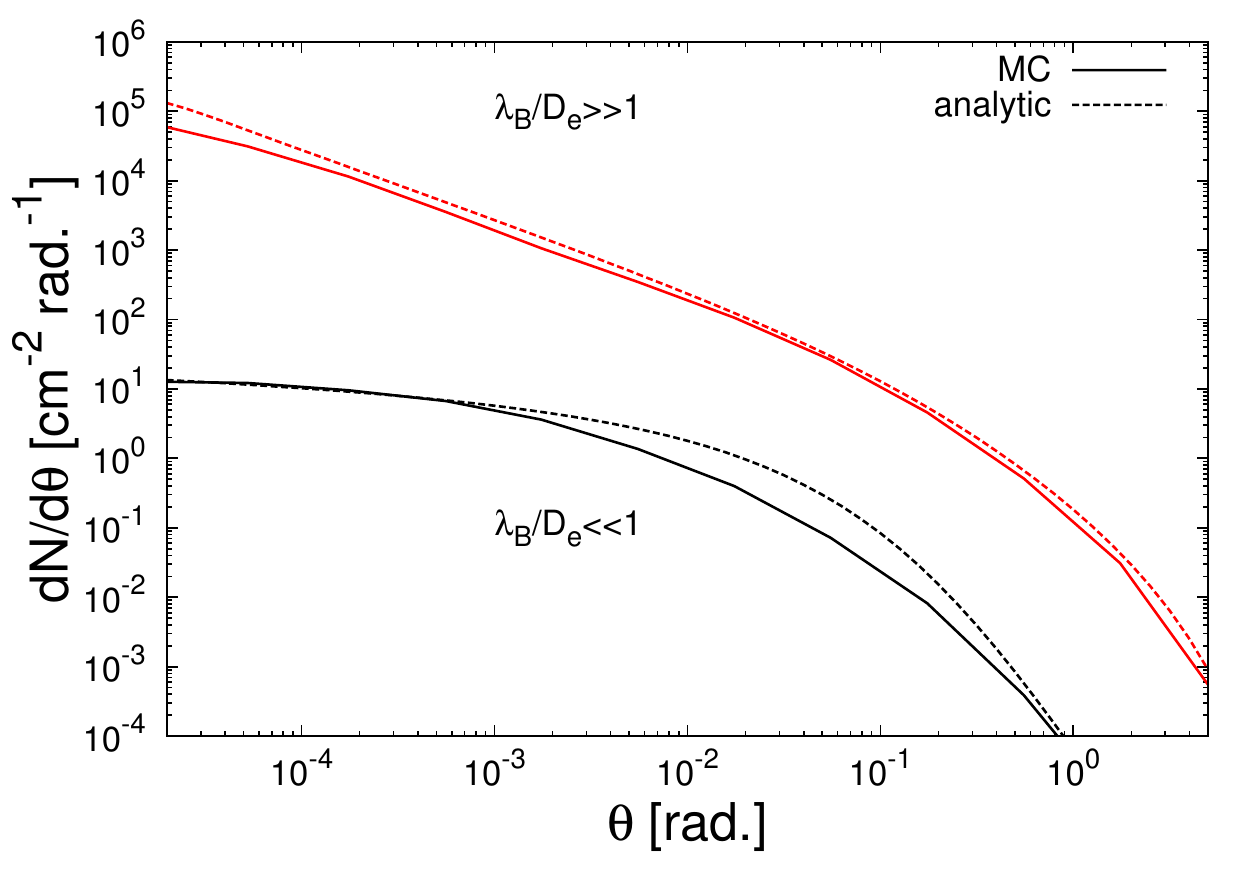}
\end{center}
\caption{The angular profile of the arriving \gr\ flux following a flaring episode
 obstained with both Monte Carlo and analytic (eqn~\ref{eq:linear1}) methods. The
angles shown are measured relative to the center of the blazar. For this plot, the 
angular profile of 1-3~GeV photons in a cascade from a source at $z=0.13$ for a $10^{-15}$G
IGMF with coherence lengths 10~kpc (black lines) and 10~Mpc (red lines) are shown.}
\label{arrival_angle}
\end{figure}

\begin{figure}[ht]
\begin{center}
\includegraphics[width=\columnwidth,angle=0]{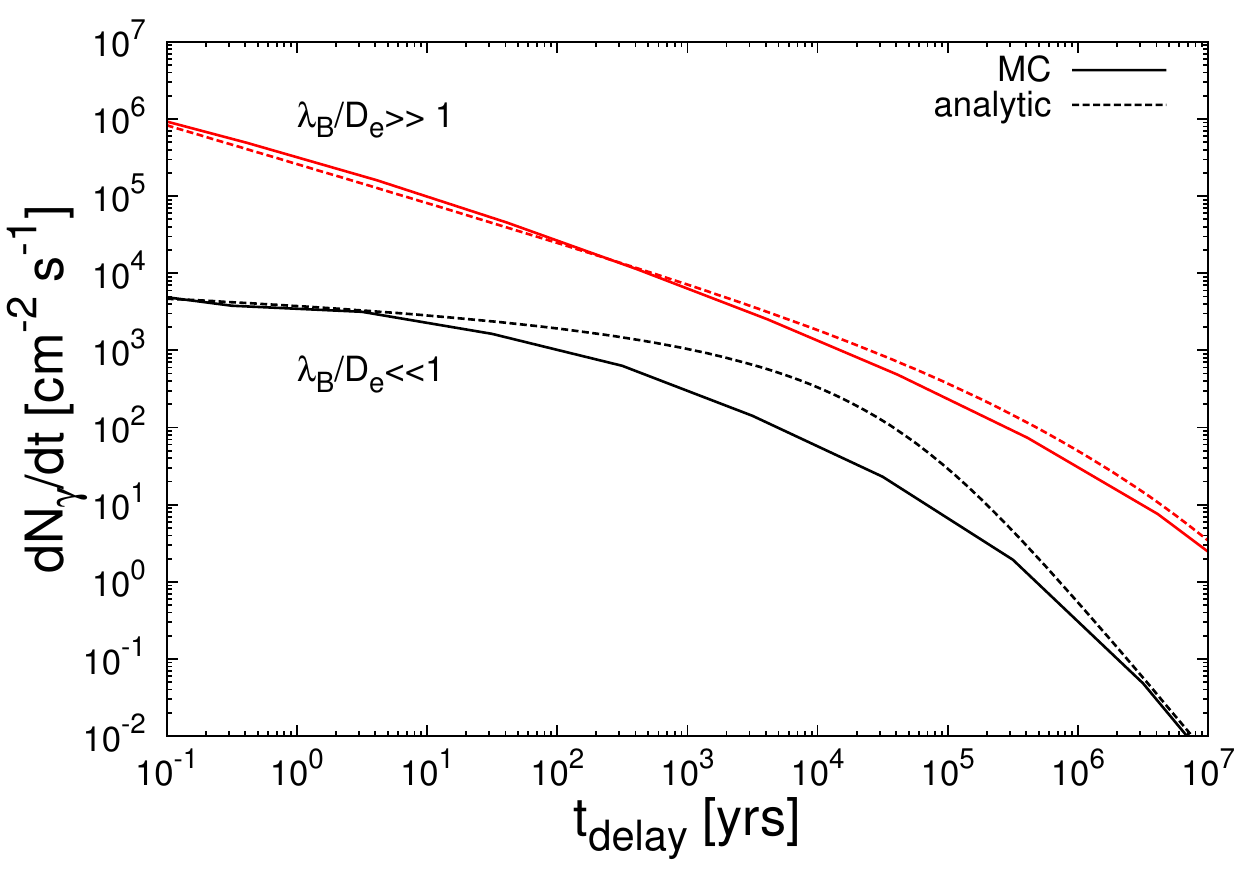}
\end{center}
\caption{The time-delay in the arriving \gr\ flux following a flaring episode
 obstained with both Monte Carlo and analytic (eqn~\ref{eq::analytic}) methods. The
time's shown are measured relative to the straight line (SO in
fig.~\ref{fig::scheme1}) arrival time. For this plot, the time-delay of
1-3~GeV photons in a cascade from a source at $z=0.13$ for a $10^{-15}$G
IGMF with coherence lengths 10~kpc (black lines) and 10~Mpc (red lines) are shown.}
\label{arrival_time}
\end{figure}

Adopting a delta-type injection spectrum with $dN/dE_{\gamma}=\delta(10^{13}~{\rm eV})$, for a blazar redshift of $z=0.13$, we compare in 
Fig.s~\ref{arrival_angle} and~\ref{arrival_time} the Monte Carlo obtained
with expressions (\ref{eq:linear1}) and (\ref{eq::analytic}).
Such comparisons confirm that the simplified analytic expressions 
obtained can indeed provide reasonably accurate descriptions of these 
distributions, particularly in the asymptotic regions.
However, we do note that some degree of divergence is found in the 
intermediate region between the asymptotic zones
for the small correlation length case ($L_{\rm B}/D_{e}\ll 1$).

\section{Conclusion}
\label{Conclusion}

Present generation \gr\ observational results have recently been
used to provide challenging new bounds on the IGMF strength. The 
coherence length for this field, however, remains largely unconstrained. 
We here consider what handle future \gr\ observations may be able to provide 
with regards a measurement of both the IGMF strength and its coherence length.
We show that measuring either the initial slope of the time delayed 
emission or the slope of the surface brightness profile of extended emission
can be used to provide a measure of the IGMF correlation length.

Through the application of a simplified analytic two-generation
model we describe two possible methods for probing this coherence length.
Though both of these methods are potentially viable, the employment of the 
first of these methods would require considerable improvement in angular resolution 
above that achieved by present day \gr\ telescopes. The second of the methods put 
forward, however, does have the potential to be applied using forthcoming \gr\ 
observational data. A subsequent comparison of the two-generation model results
with those obtained using the full Monte Carlo description confirms that
this signature is still expected to survive once the full cascade physical
description is added back into the picture.

\bibliography{AT_biblio}
\bibliographystyle{aa}

\end{document}